\documentclass[prl,amsmath,amssymb,showpacs,showkeys,twocolumn]{revtex4}
\usepackage{graphicx}
\usepackage{dcolumn}
\usepackage{bm}
\usepackage{amssymb}
\usepackage{amsmath}
\begin{document}

\title{The magnonic superfluid droplet at room temperature}

\author{Yu.~M.~Bunkov$^{(a)}$}
\email{y.bunkov@rqc.ru}
\author{A.Farhutdinov$^{(b)}$}
\author{A.~N.~Kuzmichev$^{(a)}$}
\author{T.~R.~Safin$^{(a, b)}$}
\author{P.~M.~Vetoshko$^{(a)}$}
\author{V.~I.~Belotelov$^{(a)}$}
\author{M.~S.~Tagirov$^{(b, c)}$}

\affiliation{
$^{a}$ Russian Quantum Center, Skolkovo, 143025 Moscow, Russia\\
$^{b}$ Kazan Federal University, 420008, Kazan, Russia\\
$^{c}$ Institute of Applied Research of Tatarstan Academy of Sciences, 420111, Kazan, Russia}

\date{\today}

\begin{abstract}
We declare the observation of spin superfluid state  in Yttrium Iron Garnet (YIG) at room temperature. It is similar to a Homogeneous Precessing State (HPD), observed earlier in  antiferromagnetic superfluid $^3$He-B. The formation of this state explains by the repulsive interaction between magnons, which is required as a prior condition for the spin superfluidity. It establishes an energy gap, which stabilizes the long range superfluid transport of magnetization and determines the  Ginzburg-Landau coherence length.  This discovery paves a way to many quantum applications of supermagnonics at room temperature, such as magnetic Josephson effect, long distance spin transport, Q-bit, quantum logics, magnetic sensors and others.

\end{abstract}

\pacs{67.57.Fg, 05.30.Jp, 11.00.Lm}

\keywords{Supermagnonics, spin supercurrent, magnon BEC, YIG film}

\maketitle


The coherent quantum state of matter - the superfluid state was discovered in 1938 by P. L. Kapitza in liquid $^4$He at  temperature of 2 K \cite{Kapitza}. The first theoretical explanation of this state was done by F. London. He  suggested that the superfluidity could
have some connection with Bose-Einstein condensation (BEC) \cite{London}. In BEC theory by Einstein  the coherent state of 
noninteracting particles was considered \cite{Einstein}. Recently it was successfully applied for the  weakly interacted deluted gas of atoms at extremely low temperatures \cite{aBEC,aBEC1}. But it can not be directly applied for the system with a strong inter-particle interaction, like  $^4$He liquid.  In 1941 L. D. Landau suggested that superfluidity can be understood in terms of atomic states, modified by interaction. The phenomenological theory by L. D. Landau successfully explained the superfluid properties  \cite{Landau}. The depletion of  condensate in  $^4$He  is very strong: in the limit of zero temperature only about 10\%  of particles occupy the state with zero momentum.
Nevertheless, BEC still remains the key mechanism for the phenomenon of superfluidity in liquid
$^4$He: due to BEC  the whole liquid (100\% of $^4$He atoms) forms a coherent quantum state at $T=0$ and participates in the non-dissipative superfluid flow. Nowadays the superfluid state is well known as a quantum state, governed by a single wave function. Indeed, the superfluid state may exist without BEC, like in Berezinskii–Kosterlitz–Thouless transition in 2D materials. And opposite,  BEC may exist, but does not lead to a superfluid transport. It takes place when there is no energy gap and kinetic energy destroys the coherent state. In this case the critical current is equal to zero. 

The magnetically ordered materials are described by the ground state and a gas of bosonic excitations which are represented by magnons. At a condition of thermal equilibrium  the density of  magnons is always below the critical  density, required for the  Bose condensation. Indeed, the density of magnons can be increased up to about Avagadro number by a magnetic resonance methods.  Usually the spin-spin interaction time is much shorter than the spin-lattice relaxation time and  magnon gas may exist at a quasi equilibrium state. The  magnon gas is a very interesting object to study the Bose-Einstein condensation (BEC) and formation of a spin superfluid states. Its properties are strongy depends from the type of spin-orbit interactions, which result in  attractive or repulsive interaction between magnons. At the first case the formation of magnon BEC should be unstable. Indeed, the recent experiments with magnon BEC formation shows a very interesting new dynamic effects at this case. \cite{Serga,Sergax}. Contrary, at the repulsive interaction, the magnon BEC should be stable \cite{Stamp}. Furthermore, at higher density of magnons the collective spin modes emerge. It is described by a common wave function and shows the properties comparable with the superfluid component of $^4$He liquid.  It exhibit the long distance spin supercurrent transport which is characterized by a Ginzburg-Lanadau coherence length. 

The first magnonic superfluid state was discovered in 1984 by A. S. Borovik-Romanov and Yu. M. Bunkov experimental group and described theoretically by I. Fomin as a  new state of magnetically ordered matter \cite{HPD,HPDT}. It was observed as the spontaneously self-organized phase-coherent precession of spins in an antiferromagnetic superfluid $^3$He-B.
This state fulfills all criteria of coherence, suggested later by Snoke \cite{Snoke}.
It is radically different from the conventional ordered states in magnets. It emerges on the background of the ordered magnetic state, and can be described in terms of the  condensation of magnetic excitations to a superfluid coherent quantum state \cite{MagBEC,MagBEC0,UFN}. 
The spin superfluid state exhibit many phenomena analogues to other superfluid states, including long distance spin supercurrent in the channel and phase-slippage at a critical spin supercurrent  \cite{Ssupercurrent,Ssupercurrent1,Ssupercurrent2}, Josephson
spin-current effect \cite{Joseph1,Joseph} as well as spin-current
vortices \cite{Vortex,Vortex2}. The Goldstone collective excitations of this state (the analog of the second sound in $^4$He), were also observed \cite{Goldstoun1,Goldstoun2}. The state, which is very near to a magnon BEC state, was observed in a conditions of spatial trap. In this case the BEC signal may ring for about tenth of minutes at a frequency of 1 MHz  \cite{PIS,PIS1}. This new state  can be described in a terms of a self trapping  model, suggested earlier for the formation of elementary particles \cite{Qball,Qball1}. The spin superfluid state was also observed in $^3$He-A in the conditions, when spin-orbit interaction was changed from attractive to repulsive \cite{HeABEC,HeABEC2,BVHe-A}. It was also resently observed in other  superfluid state - $^3$He-P \cite{He-P}.
The  discovery of spin superfluidity in $^3$He has been recognized in 2004  by the Low Temperature community by a prestigious F. London prize \cite{LondonPrize,LondonLecture}.

It is very important to note, that there is no any principal difference between the magnon gas in superfluid $^3$He and in solid magnetically ordered materials. The magnetic ordering is a  quantum phenomenon. The model of spatially fixed magnetic moments, usually applied at some theoretical considerations, describes only limited number of observations. For example, it fails to reproduce the temperature dependens of magnetization. The correct consideration follows from Holstein-Primakoff  transformation model \cite{HP} in which magnons - the quanta of  magnetic existations are not localized and may flow as a magnon liquid. That is why the nature of  superflow of magnetization  in superfluid $^3$He is similar to one for solid magnetic materials. The difference is only the value of Gilbert damping, which in  superfluid $^3$He can be as small as $10^{-8}$ while in solid magnetic materials it is about $10^{-5}$ in the best case  of the magnetic dielectric ferrimagnets represented by yttrium iron-garnets. Indeed, the properties of magnon flow, magnon BEC and magnon supercurrent in antiferromagnetic superfluid $^3$He and in YIG film are very similar. The first experimental observations of the spin superfluid phenomenon in a YIG film at room temperature are presented in this article.

The superfluid state is characterized by the off-diagonal long-range order (ODLRO) \cite{Yang}.
In superfluid $^4$He and in the coherent atomic systems the operators
of creation and annihilations of atoms with momentum ${\bf p}=0$ have the time-dependent  vacuum expectation value. For the creation operator
\begin{equation}
\left<\hat a_0 \right>={\mathcal N}_0^{1/2}  e^{i\mu
	t+i\alpha}\,,
\label{ODLRO}
\end{equation}
where ${\mathcal N}_0$ is the number of particles in the coherent state and $\mu$ and $\alpha$ are chemical potential and the phase of wave function, respectively.

The analogy between spin precession and ODLRO in superfluids is seen if one compares the operator of creation of particle $\hat a_0^+$  with  the operator $\hat S^+$
of creation of spin projecton on axis $z$, whose expectation value  is:
\begin{equation}
\left<\hat S^+ \right>={\mathcal S}_x +i {\mathcal S}_y ={\sqrt{{\mathcal S}-\hat{\mathcal S}_z}} \,\,\,e^{i\omega	t+i\alpha}\,.
\label{spinODLRO}
\end{equation}

This analogy suggests that in the coherent spin precession the role of  particle number  ${\mathcal N}$ is played by the projection of  total spin on the direction of the external magnetic field ${\mathcal S}_z$ \cite{Rev2,MagBEC}. The role of chemical potential is taken by the precession frequency. It is important to note, that the Zeeman energy should be incorporated to the chemical potential, since it may vary in space in the case of magnetic field inhomogeneity.  

The hydrodynamic equations for the magnon superfluid are the Hamilton equations for the canonically conjugated variables ${\mathcal N}$ and $\alpha$:
\begin{equation}
\dot\alpha= \frac{\delta F}{\delta {\mathcal N}}~~,~~ \dot {\mathcal N} =- \frac{\delta F}{\delta \alpha},
\label{HydroEq}
\end{equation}
where $F$ is the Ginzburg-Landau free energy functional of the system \cite{Book}.
This functional in a frame, rotating at a frequency $\omega$ has the conventional form
\begin{equation}
{\cal F} -\mu {\cal N}=\int d^3r\Big\{\frac{\vert\boldsymbol{\nabla}\Psi \vert^2}{2m}  + \big[\omega_L({\bf r})-\omega\big ]
\vert\Psi\vert^2 +F_{\rm so}(\vert\Psi\vert^2)\Big\}\ .
\label{GLfunctional}
\end{equation}
Here  $\omega_L({\bf r})=\gamma H({\bf r})$  is the local Larmor frequency, which plays the role of external potential $U({\bf r})$ in atomic condensates. The last term $F_{\rm so}(\vert\Psi\vert^2)$ contains nonlinearity which comes from the spin-orbit interaction. It is analogous to the 4-th order term in the atomic BEC, which describes the interaction between the atoms.  The  spin-orbit interaction provides the effective interaction between magnons, which can be attractive or repulsive.
The spin-orbit interaction contains quadratic and quartic terms in $|\Psi|$. 

\begin{equation}
F_{\rm so}(\vert\Psi\vert^2) = a \vert\Psi\vert^2 + b \vert\Psi\vert^4 + .... 
\end{equation}
While the quadratic term modifies the potential $U$ in the Ginzburg-Landau free energy, the quartic term simulates the interaction between magnons. 

We are able to rewrite Eq. \eqref{GLfunctional} in the next form
\cite{MagBEC}:
\begin{equation}\label{GLfunctional2}
{\cal F} -\mu {\cal N} = \int d^3r\Big\{\frac{\vert\nabla\Psi\vert^2}{2m} +\big[\omega_0(r) -\omega\big ]
\vert\Psi\vert^2+b \vert\Psi\vert^4)\Big\}\ ,
\end{equation}
where $\omega_0(r) = \omega_L(r) + a(r) \, \, $ is the frequency of precession at a limit of small excitation.
Here we regrope  terms with  $\vert\Psi\vert^2$ and $\vert\Psi\vert^4$. The magnon number density $N$ is related to the deflection angle $\beta$ via 
\begin{equation}
N=|\Psi|^2=M \,(1-\cos\beta\,)
\end{equation}
and the frequency of magnetization precession is
\begin{equation}
\omega_S(r) = \omega_0(r) + 2b (1-\cos(\beta(r))),
\end{equation}
where $M$ is the magnetization.

The gradients of  $\alpha$  excites the spin supercurrent, which transports the longitudinal magnetization
 \cite{Book}: 
\begin{eqnarray}
{\bf J} = {\mathcal N} \boldsymbol{\nabla}\alpha~,
\label{SpinCurrent1}
\end{eqnarray}
This long distance spin supercurrent was measured directly in the experiments, described in \cite{Ssupercurrent,Ssupercurrent1,Ssupercurrent2}. It was confirmed that the critical current corresponds to the critical phase gradient, which  is the inverse
value of the Ginzburg-Landau coherence length $\xi_{GL}$:
\begin{equation}
\nabla \alpha _c=1/\xi_{GL}=\sqrt{\omega_0 (\omega_{S}-\omega_0)}
/c_{SW},
\end{equation}
where $c_{SW}$ is a spin wave velocity. It is determined by the competition between the energy of repulsive interaction (third term in Eq. 6) and kinetic energy of flow (first term in Eq. 6). Please note, that at the attractive interaction the coefficient $b$ is negative and spin supercurrents unstable.

The remarkable consequence of the spin superfluidity is the formation of the homogeneous precession domain (HPD) even in a strongly inhomogeneous magnetic field. It is formed because the gradient of magnetic field leads to a gradient of phase of magnetization precession which excites the spin supercurrent. The latter transports magnons to the direction of smaller field. The precession frequency increases untill the gradient of precession vanishes. Finally the equilibrium state with the precession frequency  $\, \omega_S(r) = const \,$ is established. Any perturbation of this state excites the spin supercurrent which restore the coherent precession.  It was shown that the HPD state appears spontaneously at some delay after the pulse of magnetic resonance excitation.  This state radiates an extremely long living induction signal \cite{HPD,HPDT}.
It can persist for several minutes due to the very slow relaxation (evaporation) of quasiparticles \cite{Persist3,Qball1}. Recently this type of the signal was considered as a time-crystal \cite{Tcrystal}.
It is important to note that the HPD state is the eigen state of the ensemble of excited magnons.
At  relaxation the number of magnons decreases, the frequency of HPD decreases but the magnons  remain in the coherent state.

The other distinctive feature of the HPD state is its permanence. In the case of the atomic BEC the state disappears because of the atom evaporation. In the case of HPD it is possible to replenish the losses (evaporation) of quasiparticles by  excitation of  new quasiparticles. It was shown experimentally \cite{CWBEC,HeABEC,spin}, that a weak RF pumping at a frequency $\omega_{S}$ stabilizes chemical potential of magnon gas and keeps its density constant. The corresponding phase difference between the magnetization precession and the RF field appears automatically to compensate the energy losses.

In this article we describe the first observation of HPD state in an out-of-plane magnetized YIG film. 
The experiments were performed on a YIG films of 6 and 1 $\mu$m thickness in a shape of a disks 0.5 and 0.3 mm in diameter (see Methods).

The magnetization precession in the out-of-plane magnetized YIG film has dynamic properties very similar to ones in a superfluid $^3$He-B. The repulsive interaction between magnons leads to the dynamical frequency shift of the precession at its deflection on angle $\beta$ \cite{Gulaev2000} described by equation:
\begin{equation}
\omega - \omega_0=  \gamma 4\pi M{_S}(1-\cos{\beta}),
\label{freqshiftNoRF}
\end{equation}
where $\omega$ is magnon frequency of the homogeneous precession (k=0) at the deflection angle $\beta$, $\omega_0$ is magnon frequency at the limit of small excitation, $\gamma$ is gyromagnetic ratio, and $M{_S}$ is saturation magnetization. It means that the local frequency grows with increasing magnon density. As a result the resonance field decreases if we excite the system at a constant frequency.

We have investigated the FMR adsorption signals from the YIG film at frequency 9.26 GHz and different RF power of excitation. The signals are shown in Fig. 1. At first sight it looks like a well known signals of the non-linear (bi-stable) resonance, first described by Anderson et al. \cite{Anderson}. However  this theory does not correspond well to the experimental results at relatively high excitation \cite{Fetisov1999}. The reason for this discrepancy is that the Anderson's theory doesn't take into account the spin supercurrents, which
play a very important role at a high angles of the  magnetization deflection, when the density of magnons surpase the conditions of magnon Bose-Einstein condensation, which is about 2-3 $^\circ$ for YIG film \cite{Bunkov2018a}. 

At a relatively small  excitation power of 0.05, 0.1 and 0.4 mW (see Fig. 1B) the amplitude of   absorption signal grows proportionally to the RF field. This property is correspond to an excitation of the magnon  gas. At higher excitation the ``capture'' of a signal takes place. The signals start to follow the field change (see curve c in Fig. 1B).  This non-linear behavior starts at an angle of deflection about 2-3$^\circ$, when the field shift of the resonance became bigger then the broadening of the resonance line. This angle of deflection is corresponds to the theoretical approximation for magnon BEC formation \cite{Bunkov2018a}. At higher excitation the signal follows the magnetic field up to its change by
80 Oe, as follows from the curve d at the Fig. 1A. This field shift corresponds to a frequency shift of about  320 MHz and the angle of magnetization deflection (see Eq. 7) by about 20$^\circ$! 

At some critical field change the signal disintegrate.  This critical  field shift (CFS) strongly depends on the exciting power. If we sweep the field up the signal restore at a some field, which is quite near to the field of disintegration (see dotted line d in Fig. 1A). The observed behavior of the signal amplitude shows that the signals are generated by a state with coherent precession of magnetization, which appears due to a spatial magnons redistribution by the spin supercurrent. Its properties are wery similar to one in antiferromagnetic superfluid $^3$He, and particularly to the signals in $^3$He-A in aerogel \cite{HeABEC,HeABEC2,HeAteor}.

\begin{figure}[htt]
	\includegraphics[width=0.5\textwidth]{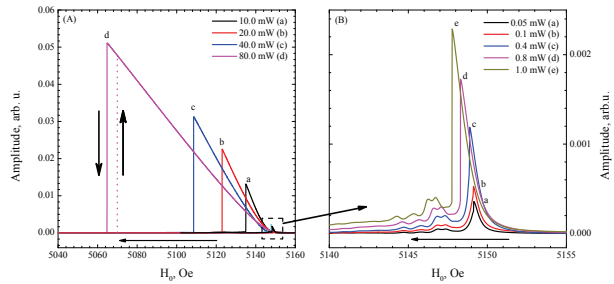}
	\caption{Amplitudes of the absorption signals at a frequency of 9.26 GHz for different RF pumping powers at a magnetic field sweep down. The curves a, b and c in Fig. B corresponds to a linear magnetic resonance, when the amplitude of the signal is proportional to the RF field. The curve c correspond to conditions, when the magnon BEC should forms. Curves at higher excitation correspond to the conditions to a non-linear behavior of signals when the magnon superflow plays an important role. It redistribute the magnetization in the sample and provide the stability of coherent precession. Curve d in Fig A corresponds to a highest energy of excitation at this experiment.  The field shift of the signal about 80 Oe was achieved. It corresponds to an angle of precessing magnetization deflection of about 20$^\circ$. The curve d, which shown by dotted line corresponds to a signal we have observed at a sweep field up. It shows that the signal of non-linear magnetic resonance restores at a small hysteresis.
	(The power of 100 mW in these experiments  corresponds roughly to the RF magnetic field of 0.1 Oe.) }
	\label{CWsmallP}
\end{figure}

According to a theoretical consideration for the out-of-plane magnetized yttrium iron garnet film the density of excited magnons should reach the conditions of BEC at the angle of magnetization deflection  about  $\beta = 2,5^\circ$ \cite{Bunkov2018a}. This angle well corresponds to a begining of signals nonlinearity (see Fig. 1). 
At an excitation power higher then 0.4 mW for the conditions of our experiments the density of magnons increases and  strong repulsive interaction between magnons  modifies the spectrum of magnons. The spin dynamics can not be described anymore by the BEC theory of weekly interacted magnons. The formation of coherent macroscopic states and long distance spin supercurrent should be considered.

\begin{figure}[htt]
\includegraphics[width=0.4\textwidth]{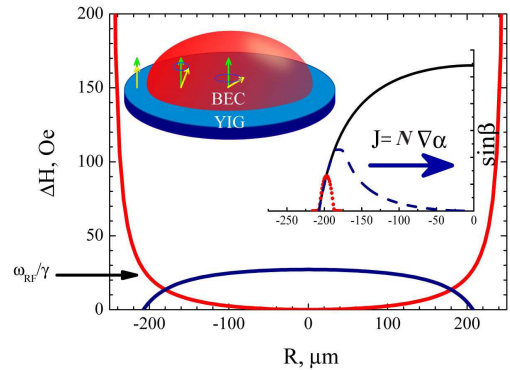}
\caption{Variation of the local effective magnetic field from the center of the sample (red line) and the frequency shift due to the magnon chemical potential (blue line), which compensates the inhomogeneity of magnetic field. The inset to the right shows schematically the spatial distribution of  magnetization deflection ($\sin \beta$)  at small excitation (point red line), at a higher excitation, when magnons start to superflow to lower field parts of the disk (dashed blue line) and the HPD state (solid black line). The inset to the left side shows the HPD droplet of coherent magnons. }
	\label{field}
\end{figure}

In Fig. 2  the profile of effective magnetic field in the 500 $\mu$m in diameter disk-shaped YIG film is shown. It was calculated by OOMMF micromagnetic software \cite{66}. The effective field has a minimum at the disk center due to the demagnetization field. One may suggest that at a field sweep down  the resonance conditions appear first at a central part. Then the magnetization deflects and the frequency follows to the resonance conditions at a smaller field untill the signal breaks down. But why the signal restore at a field sweep up with a very small hysteresis. Let us consider the case, when the resonanse conditions appears only near the edge of the sample.  
If we apply the RF field at the frequency corresponding to the ferromagnetic resonance near the edge (at R = 200 $\mu$m in Fig. 2) the spin waves will be excited at this region (see the right inset Fig. 2 red dot line). If the excitation power will be increased the deflection angle also increases and may surpasses the critical BEC angle. The gradient of  phase of precession and spin supercurrent will appears due to the gradient of effective magnetic field. It will transport magnons in the direction of smaller field, to the central part of the sample (Fig. 2 dashed blue line). Finally, the angle of magnetization deflection in the central part increases up to a value, which corresponds to the precession frequency equal to RF field (black line). The spatial distribution of dynamic frequency shift is shown in Fig. 2 by a solid blue line. This shift compensates the inhomogeneity of effective magnetic field and all the magnetization precesses at the frequency of RF field.  The HPD state is the eigenstate in inhomogeneous magnetic field for a given number of magnons. This scenario was carefully investigated in the experiments with superfluid $^3$He and MnCO$_3$. \cite{CWBEC,3MnCO3}.   
 
 The HPD state has a form of droplet, artistically shown in the left inset in Fig. 2.
Deflected magnetization precesses coherently and homogeneously at a frequency:
\begin{equation}
\omega = \gamma H(r) + \Delta \omega (r) = const,
\end{equation}
where $H(r)$ is the effective magnetic field at a point $r$.  Furthermore, the spin supercurrent compensates the inhomogeneity of magnetization relaxation by redistribution of magnons density.
This state exists permanently in the case of RF pumping at the frequency of the HPD state, which compensates the evaporation of magnons.

\begin{figure}[htt]
 \includegraphics[width=0.45\textwidth]{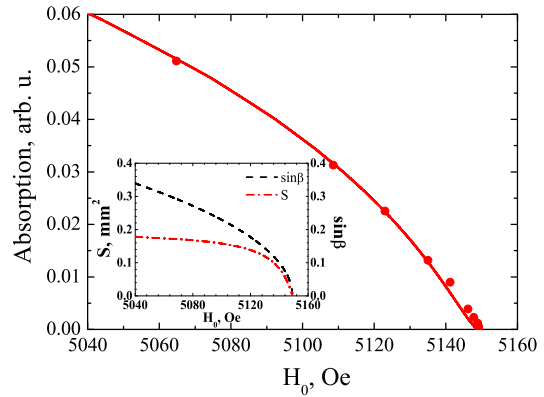}
 \caption{The absorption signal amplitude at the moment of breaks down of the HPD state (points). The  theoretical curve (red solid line) was calculated for the signal from the magnonic droplet at these conditions.  The inset shows the magnetization deflection ($\sin \beta$) in the center of the magnon droplet and the film surface area, occupied by the droplet (S). }
 \label{Ampl}
\end{figure}

The superfluid state is supported by a permanent pumping of the RF magnetic field, ${\bf H}_{\rm RF}$, which is
transverse to the applied constant external magnetic field ${\bf H_0}$.   The RF field prescribes the
precession frequency, $\omega=\omega_{\rm RF}$, and thus fixes the chemical
potential $\mu= \omega$. In the precession frame, where both the RF field
and the deflected magnetization  ${\bf M}$ are constant,  the interaction energy term is
\begin{equation}
F_{\rm RF}=- {\bf H}_{\rm RF}\cdot{\bf M}= -   H_{\rm
RF} M_\perp \cos(\alpha-\alpha_{\rm RF}) ~, \label{InteractionRF}
\end{equation}
where $H_{\rm RF}$ and $\alpha_{\rm RF}$ are the amplitude and the phase of
the RF field. This term  softly
breaks the $U(1)$-symmetry and serves as a source of the mass of Nambu-Goldstone mode
\cite{Volovik2008}.
The phase difference between the condensate and  RF field, ($\alpha-\alpha_{RF}$), is determined by the
energy losses due to magnetic relaxation, which is compensated by
the pumping  power of the RF field:
\begin{equation}
  W = \omega MH_{\rm RF}\sin\beta \sin{(\alpha-\alpha_{\rm RF})}.
  \label{pumping}
\end{equation}
The phase difference  is
automatically adjusted to compensate the losses. If dissipation is
small, the phase shift is small ($\alpha-\alpha_{\rm RF}\ll 1$) and can be neglected.
The neglected $(\alpha-\alpha_{\rm RF})^2$ term leads to the nonzero mass of the
Goldstone boson -- quantum of the second sound waves (phonons)
in the magnonic superfluid \cite{Volovik2008}.
The signal  breaks down at the moment, when the RF power is not enough to compensate the magnons dissipation.
Since the pumping \eqref{pumping} is proportional to $\sin\beta\sin(\alpha-\alpha_{\rm RF})$,
a critical tilting angle $\beta_c$, at which the pumping cannot
compensate the losses,  increases with increasing $H_{RF}$. The breaks down
occurs when the phase shift ($\alpha-\alpha_{\rm RF}$) reaches  90$^\circ$.
At this moment the adsorption signal corresponds to a transverse magnetization of the sample.
 In Fig. \ref{Ampl} it is shown by points. 
The theoretical curve, shown in Fig. \ref{Ampl} by solid line,  is calculated for the assumption that the HPD droplet fills up all the region, where magnetic field is less than $\omega_{\rm RF}/\gamma$.  The theoretical curve shows a good agreement with the experimental points. The similar results were obtained in other samples.

Let us compare the experimental results with the theory of non-linear resonance  \cite{Anderson}. This theoretical approach is based on the properties of a single non-linear oscillator. It supposes that at increase of the excitation power, the angle of deflection increases which leads to a  frequency shift. This theory describes well the non-linear resonance in ferromagnets at a relatively small angles of deflection but strongy disagrees with the experimental results at higher angles \cite{Fetisov1999}. A more sophisticated  theory of autoresonance was suggested in \cite{Autoresonance}, where a singe non-linear oscillator is also considered. This model is not applicable to real magnetically ordered materials since the spin system consists of many oscillators with different ground frequency due to inhomogeneity of magnetic field. If we try to modify the theory for a number of the non-interacting oscillators with different ground frequencies  we should  conclude that the break down points are different for different oscillators. As a result the experimental curve of signal break down should be broaden on the magnetic field. Indeed, experimentally the break down appears very abruptly. The even more important argument against this theory is the recovery of  signal with a very small hysteresis at a sweep magnetic field up. The non-liner oscillator can not be excited by out of resonance excitation. It is clearly shown in Fig. 3 of Ref.  \cite{Autoresonance}. It may be excited at a sweep field up only near the resonance, at a field shift of  a homogeneous broadening of the line, which is  about 2 Oe. Experimentally the excitation takes place at uncomparable big shift of 80 Oe. This can be explained only by a resonance  excition of magnons at the edge and its  redistribution  throw all the sample by spin supercurrent. The comparison between the model of  non-linear oscillator and HPD have been investigated in Ref. \cite{Okinava}.

It is very important to note that the spin supercurrent exists only at the conditions of repulsive interaction between magnons. In the case of attractive interaction the superfluid critical velocity is equal to zero and the superfluid state is unstable \cite{Unstable}. 

Finally, we have shown the big importance of spin supercurrent transport  to explain the properties of the non-linear magnetic resonance. We have demonstrated the formation of a macroscopic region with the coherent precession of magnetization in an inhomogeneous effective magnetic field in the out-of-plane magnetized YIG film.  This state is the first permanent superfluid state of condensed matter demonstrated at room temperature. It is an ideal platform for  development of microwave magnetic technologies, which have already resulted in the creation of  the magnon transistor and the first magnon logic gate \cite{Serga2017,Kajiwara2010}.
The YIG films can be used as the basis for new  solid-state quantum measurement and information processing technologies including cavity-based QED, optomagnonics, and optomechanics \cite{Zhang2015}.
A chain of YIG samples with excited HPD droplets may be considered as a Q-bits, interacting through Josephson junctions for  a quantum computer. The formation of the magnon BEC in YIG and  observation of the spin supercurrent, like in $^3$He, should lead to the development of a new branch of modern magnetism - supermagnonics.

\subsection{ Methods}
All samples  were prepared from yttrium iron garnet films grown by liquid phase epitaxy on 500 $\mu$m thick GGG substrates with the (111) crystallographic orientation \cite{sample}. To reduce the effect of cubic magnetic anisotropy, we used scandium — substituted $Lu_{1.5}Y_{1.5}Fe_{4.4}Sc_{0.6}O_{12}$ iron garnet films; the introduction of lutetium ions was necessary to match the parameters of the substrate and film crystal gratings. It is known that the introduction of scandium ions in such an amount reduces the field of cubic anisotropy by more than an order of magnitude \cite{Sc}. In addition, the used lutetium and scandium ions practically do not contribute to additional relaxation in the YIG.
The samples were prepared in the form of a disk with diameters of 500 and 300 $\mu$m and a thickness of 6 $\mu$m.  The disk was made by photolithography. To avoid magnetic pinning on the surface the sample was etched in a hot phosphoric acid \cite{Vet}. As a result, the edges of the disk have a slope of 45 degrees and had a smooth surface.

The CW FMR experiments were performed on Varian E-12 X-band EPR spectrometer at the room temperature and the frequency 9.26 GHz.  The RF field  was oriented in plane of the samples. The amplitude and the frequency of magnetic field modulation were 0.05 Oe and 100 kHz, respectively. This frequency is much lower than the estimated frequency of the second sound of the magnon BEC (The Goldston mode). That is why we may consider these conditions as stationary. The absorption signals, presented here, were obtained after the integration of the original signals.

\subsection{ Acknowledgments}
\acknowledgments The authors wish to thank G. E. Volovik, V. P. Mineev  V. L'vov and A. Serga for helpful
comments and O. Demokritov for stimulating discussions. Financial support by the Russian Science Foundation within the Grant 19-12-00397 “Spin Superfluids” is gratefully acknowledged.


\begin{thebibliography}{59}
	
\bibitem{Kapitza} P	Kapitza,   ``Viscosity of liquid helium below the $\lambda$-point'' {\it Nature} {\bf 3558},  74 (1938).

\bibitem{London} F. London, {\it ``Superfluids''}, {\bf II} John Wilet and Sons, Inc., New York, (1954).

\bibitem{Einstein} A. Einstein, 
 "Quantentheorie des einatomigen idealen Gases. Part I".
{\it Sber. Preuss. Akad. Wiss.}, {\bf 22,} 261 (1924); 
"Quantentheorie des einatomigen idealen Gases. Part II". {\it Sber. Preuss. Akad. Wiss.}
 {\bf 1,} 3, (1925).

\bibitem{aBEC} M. H. Anderson, J. R. Ensher,  M. R. Matthews,  C. E. Wieman \&  E. A. Cornell, ``Observation of Bose-Einstein condensation in a dilute atomic vapor", 
{\it Science} {\bf 269,} 198 (1995).

\bibitem{aBEC1} K. B.  Davis, M. O. Mewes, M. R. Andrewes, N. J. van Druten, D. S. Durfee, D. M. Kurn, W. Ketterle, 
`` Bose-Einstein condensation in a gas of sodium atoms'',
 {\it Phys. Rev. Lett.}, {\bf 75}, 3969 (1995).

\bibitem{Landau} L. D. Landau, 
``Theory of helium 2 superfluidity'' 
 {\it J. Phys. USSR}, {\bf 5}, p.71 (1941).

\bibitem{Serga} D. A. Bozhko, A. A. Serga, A. Pomyalov, V. S. L’vov  and B.  Hillebrands,
``Bogoliubov waves and distant transport of magnon condensate at room temperature''.
{\it Nature Comm.} {\bf 10}, 2460 (2019).


\bibitem{Sergax} O. Dzyapko, I. Lisenkov, P. Nowik-Boltyk, V. E. Demidov, S. O. Demokritov,
B. Koene, A. Kirilyuk, T. Rasing, V. Tiberkevich and A. Slavin,
``Magnon-magnon interactions in a room-temperature magnonic Bose-Einstein condensate".
{\it Phys. Rev. B} {\bf 96}, 064438 (2017).

\bibitem{Stamp} I. S. Tupitsyn, P. C. E. Stamp, A. L. Burin, 
``Stability of Bose-Einstein condensates of hot magnons in YIG film'', 
 {\it Phys. Rev. Lett.} {\bf 100}, 257202 (2008).

\bibitem{HPD} A. S. Borovik-Romanov, Yu. M. Bunkov, V. V. Dmitriev and Yu. M. Mukharskii,
``Long-lived induction signal in superfluid $^{3}$He-B".
{\it JETP Lett.} {\bf 40,} 1033 (1984).

\bibitem{HPDT} I. A. Fomin,
``Long-lived induction signal and spatially nonuniform spin precession in $^{3}$He-B".
{\it JETP Lett.} {\bf 40,} 1037 (1984).


\bibitem{Snoke} D. Snoke,
"Coherent questions".
{\it Nature} {\bf 443,} 403 (2006).

\bibitem{MagBEC0} Yu. M. Bunkov 
``Spin Supercurrent''
 {\it J. Mag. Mag. Mat}, {\bf 310},  1476 (2007).


\bibitem{MagBEC} Yu. M. Bunkov and G. E. Volovik,
``Bose-Einstein Condensation of Magnons in Superfluid $^{3}$He".
{\it J. of Low Temp. Phys.} {\bf 150,} 135 (2008).



\bibitem{UFN}    Yu. M. Bunkov,
``Spin superfluidity and magnons Bose–Einstein condensation''
{\it Physics Uspekhi,} {\bf  53,} 848 (2010).

\bibitem{Ssupercurrent} Yu. M. Bunkov,  
``Spin Supercurrent in 3He-B'', 
{\it Japan 	J. Appl. Phys}, {\bf 26}, 1809 (1987).

\bibitem{Ssupercurrent1} A. S. Borovik-Romanov, Yu. M. Bunkov, V. V. Dmitriev and  Yu. M. Mukharskiy,
``Phase Slipage Observations of Spin Supercarrent in 3He-B'',
{\it JETPh Lett.,}  {\bf 45}, 124 (1987).

\bibitem{Ssupercurrent2} A. S. Borovik-Romanov, Yu. M. Bunkov, V. V. Dmitriev, Yu. M. Mukharskiy and D. A. Sergatskov,
``Investigation of Spin Supercurrent in 3He-B'',
{\it Phys.Rev.Lett.} {\bf 62}, 1631 (1989).

\bibitem{Joseph1} A. S. Borovik-Romanov, Yu. M. Bunkov, V. V. Dmitriev, V. Makroczyova, Yu. M. Mukharskii, D. A. Sergatskov, A. de Waard, 
 `` The analog of the Josephson  Effect in the Spin Supercurrent'' 
{\it Journal de Physique} {\bf 49 (C8)} 2067 (1988).

\bibitem{Joseph} A. S. Borovik-Romanov, Yu. M. Bunkov, A. de Waard, V. V. Dmitriev,
V. Makrotsieva, Yu. M. Mukharskiy and D. A. Sergatskov,
 ``Observation of a Spin Supercarrent Analog of the Josephson  Effec'',
{\it JETP Lett.,} {\bf 47}, 478 (1988).

\bibitem{Vortex} A. S. Borovik-Romanov, Yu. M. Bunkov, V. V. Dmitriev, Yu. M. Mukharskiy and
D. A. Sergatskov,
 ``Observation of Vortex-like Spin Supercurrent in 3He-B'',
{\it Physica B,} {\bf 165}, 649 (1990).

\bibitem{Vortex2} Yu. M. Bunkov and G. E. Volovik
``Spin Vortex in magnon BEC of Superfluid 3He-B'',
{\it Physica C,} {\bf 468}, 600 (2008).


\bibitem{Goldstoun1} Yu. M. Bunkov, V. V. Dmitriev, and Yu. M. Mukharskii,
``Torsional vibrations of a domain with uniform magnetization precession in $^{3}$He-B".
{\it JETP Lett.} {\bf 43,} 131-134 (1986).

\bibitem{Goldstoun2}  Yu. M. Bunkov, V. V. Dmitriev, and Yu. M. Mukharskii,
``Low frequency oscillations of the homogeneously precessing domain in $^{3}$He-B".
{\it Physica B} {\bf 178,} 196 (1992).

\bibitem{PIS} Yu. M. Bunkov, S. N. Fisher, A .M .Guenault,  G. R .Pickett, 
``Persistent Spin Precession in 3He-B in the Regime of Vanishing Quasiparticle Density'',  
{\it Phys. Rev. Lett.}, {\bf 69}, 3092 (1992).


\bibitem{PIS1} Yu. M. Bunkov, 
``Persistent Signal; Coherent NMR state Trapped by Orbital Texture''  
{\it J. Low Temp. Phys}, {\bf 138}, 753 (2005).

\bibitem{Qball} Yu. M. Bunkov and G. E. Volovik, 
``Magnon Condensation into a Q Ball in 3He-B''
{\it Phys. Rev. Lett.}, {\bf 98}, 265302 (2007).


\bibitem{Qball1} S. Autti, Yu. M. Bunkov, V. B. Eltsov,  et al.
“Self-trapping of magnon Bose-Einstein condensates in the ground and excited levels: from harmonic  to a box confinement”  
{\it Phys. Rev. Lett.}, {\bf 108}, 145303 (2012).

\bibitem{HeABEC} T. Sato, T. Kunimatsu, K. Izumina, A. Matsubara, M. Kubota, T. Mizusaki, and Yu. M. Bunkov
``Coherent Precession of Magnetization in the Superfluid 3He A-Phase''
{\it Phys. Rev. Lett.}, {\bf 101}, 055301 (2008).

\bibitem{HeABEC2} P. Hunger, Y. M. Bunkov, E. Collin and H. Godfrin
``Evidence for Magnon BEC in Superfluid 3He-A''
{\it J.  Low Temp. Phys.} {\bf 158}, 129 (2010).


\bibitem{BVHe-A} Yu. M. Bunkov, and G. E. Volovik,
``Magnon BEC in superfluid $^{3}$He-A".
{\it JETP Lett.} {\bf 89}, 306-310 (2009)

\bibitem{He-P} S. Autti, V. V. Dmitriev, J. T .Makinen, T. Rysti, A. A. Soldatov, G. E. Volovik, A. N. Yudin, V. B. Eltsov, 
``Bose-Einstein Condensation of Magnons and Spin Superfluidity in the Polar Phase of $^3$He'',
{\it Phys.Rev.Lett.} {\bf 121}, 025303 (2018).



\bibitem{LondonPrize}  
``The 2008 Fritz London Prize''	
{\it J.  Low Temp. Phys.} {\bf 152,} 1 (2008).

\bibitem{LondonLecture} Yu. M. Bunkov, 
``Spin Supercurrent and coherent spin precession'' London prize lecture,
{\it J. Phys.:Cond. Mat.}, {\bf 21}, 164201 (2009).

\bibitem{HP} T. Holstein and H. Primakoff,
``Field Dependence of the Intrinsic Domain Magnetization of a Ferromagnet'' 
{\it Phys. Rev.} {\bf58,} 1098 (1940).


\bibitem{Yang}
C.N. Yang,
``Concept of off-diagonal long-range order and the quantum phases of liquid He and of superconductors'',
{\it Rev. Mod. Phys.}, {\bf 34}, 694 (1962).

\bibitem{Rev2} Yu. M. Bunkov and  G. E. Volovik,
``Magnon Bose-Einstein condensation and spin superfluidity".
{\it J. Phys.: Cond. Mat.} {\bf 22,} 164210 (2010).



\bibitem{Book} Yu. M. Bunkov and G. E. Volovik, {\it ``Spin Superfluidity and Magnon BEC''} in {\it Novel Superfluids Ch.4,} (eds. Bennemann, K. H. \& Ketterson, J. B. Oxford Univ. Press, Oxford, (2013).


\bibitem{Persist3}
D. J. Cousins, S. N. Fisher, A. I. Gregory, G. R. Pickett and N. S. Shaw,
``Persistent coherent spin precession in superfluid $^3$He-B driven by off-resonant excitation'',
{\it Phys. Rev. Lett.},  {\bf 82}, 4484 (1999).




\bibitem{Tcrystal} S. Autti, V. B. Eltsov, and G. E. Volovik,
``Observation of a time quasicrystal and its transition to a super fluid time crystal'',
{\it Phys. Rev. Letts.} {\bf 120}, 215301 (2018).



\bibitem{CWBEC} A. S. Borovik-Romanov, Yu. M. Bunkov, V. V. Dmitriev, Yu. M. Mukharskii,
E. V. Poddyakova and O. D. Timofeevskaya,
``Distinctive Features of a CW NMR in $^{3}$He-B due to a Spin Supercurrent''.
{\it JETP}, {\bf 69,} 542 (1989).


\bibitem{spin} Yu. M. Bunkov,
``Magnonics and Supermagnonics''
{\it Spin}, {\bf 9}, 1940005 (2019).

\bibitem{Gulaev2000}  Yu. V. Gulyaev, P. E. Zilberman, A. G. Temiryazev and M. P. Tikhomirova,
``Principal Mode of the Nonlinear Spin-Wave Resonance in Perpendicular Magnetized Ferrite Films".
{\it Physics of the Solid State} {\bf 42,} 1062 (2000).


\bibitem{Anderson} P. W. Anderson and H. Suhl, 
``Instability in the motion of ferromagnets at high microwave power levels'', 
{\it Phys.	Rev.}, {\bf 100}, 1788 (1955).

\bibitem{Fetisov1999} Yu. K. Fetisov, C. E. Patton and  V. T. Synogach,  
``Nonlinear Ferromagnetic Resonance and Foldover in Yttrium Iron Garnet Thin Films—Inadequacy of the Classical Model", 
{\it IEEE Transactions on magnetics}, {\bf 35,} 4511 (1999).


\bibitem{Bunkov2018a} Yu. M. Bunkov and V. L. Safonov,
``Magnon condensation and spin superfluidity".
{\it J. Mag. and Mag. Mat.} {\bf 452,} 30 (2018).

\bibitem{HeAteor} Bunkov, Yu. M. and Volovik, G. E. 
"On the possibility of the Homogeneously Precessing Domain in Bulk $^{3}$He-A", 
{\it Europhys. Lett.}, {\bf 21,} 837 (1993).


\bibitem{66} https://math.nist.gov/oommf/software.html



\bibitem{3MnCO3} Y. M. Bunkov, A. V. Klochkov, T. R. Safin, K. R. Safiullin and M. S. Tagirov,
``Nonresonance excitation of magnon Bose-Einstain  condensation in  MnCO$_3$'',
{\it JETP. Lett.}, {\bf 109}, 43 (2019).



\bibitem{Volovik2008} G.E. Volovik,
``Phonons in magnon superfluid and symmetry breaking field",
{\it JETP Lett.}, {\bf 87,} 639 (2008)


\bibitem{Autoresonance} M. A. Shamsutdinov, L. A. Kalyakin and A. T. Kharisov, 
``Autoresonance in a Ferromagnetic Film'',
 {\it Technical Physics}, {\bf 55}, 860 (2010).

\bibitem{Okinava} L. V. Abdurakhimov, M. A. Borich, Yu. M. Bunkov, R. R. Gazizulin,
D. Konstantinov, M. I. Kurkin and A. P. Tankeyev, 
``Nonlinear NMR and magnon BEC in antiferromagnetic materials with coupled electron and nuclear spin precession'' 
 {\it Phys. Rev. B}, {\bf 97}, 024425 (2018)

\bibitem{Unstable} A. S. Borovik-Romanov, Yu. M. Bunkov, V. V. Dmitriev, Yu. M. Mukharskiy
``Instability of Homogeneous  Spin Precession  in Superfluid 3He-A'',
 {\it JETP Lett.}, {\bf 39}, 469 (1984).


\bibitem{Serga2017} A. V. Chumak, A. A. Serga and B.Hillebrands,
``Magnonic crystals for data processing''.
{\it J. Phys. D: Appl. Phys.}, {\bf 50,} 244001 (2017). 

\bibitem{Kajiwara2010} Y. Kajiwara, K. Harii, S. Takahashi, J. Ohe, K. Uchida, M. Mizuguchi, H. Umezawa, H. Kawai, K. Ando, K. Takanashi, S. Maekawa and E. Saitoh
``Transmission of electrical signals by spin-wave interconversion in a magnetic insulator",
{\it Nature}, {\bf 464,} 262-266 (2010).


\bibitem{Zhang2015} Dengke Zhang, Xin-Ming Wang, Tie-Fu Li, Xiao-Qing Luo, Weidong Wu, Franco Nori and J. Q. You
``Cavity quantum electrodynamics with ferromagnetic magnons in a small yttrium-iron-garnet sphere".
{\it NPJ Quant. Inf.}, {\bf 1,} 15014 (2015).

\bibitem{sample} The samples were  provided by the company ``M-Granat" (http://m-granat.ru/).

\bibitem{Sc} A. R. Prokopov, P. M. Vetoshko, A. G. Shumilov, A. N. Shaposhnikov, A. N.
Kuz'michev, N. N. Koshlyakova, V. N. Berzhansky, A. K. Zvezdin and V. I. Belotelov,
“Epitaxial BiGdSc iron-garnet films for magnetophotonic applications,”
{\it J. Alloys and Compounds}, {\bf 671}, 403 (2016).

\bibitem{Vet} P. M. Vetoshko, A. K. Zvezdina, V. A. Skidanov, I. I. Syvorotka,
I. M. Syvorotka and V. I. Belotelov,
``The Effect of the Disk Magnetic Element Profile on the Saturation Field and Noise of a Magneto-Modulation Magnetic Field Sensor''.
{\it Technical Phys. Lett.}, {\bf 41,} 458 (2015).




\end{thebibliography}
\end{document}